\documentstyle[12pt]{article}
\newcounter{mycount}

\newcommand{\be}{\begin{eqnarray}}
\newcommand{\ee}{\end{eqnarray}}

\newcommand\half{\frac 1 2 }

\newcommand\noi{\noindent}

\begin{document}
\bibliographystyle{nphys}
\noindent P.~N.~Lebedev Institute Preprint     \hfill
FIAN/TD/17--93\\ I.~E.~Tamm Theory Department       \hfill
\begin{flushright}{August 1993}\end{flushright}
\ \ \ \ \
\medskip
\phantom{MMMMMMMMMM} \\
\centerline{\Large\bf Matrix Versions of the Calogero Model}

\vskip 10 mm

\centerline{O.V.~Dodlov,  S.E.~Konstein, and M.A.~Vasiliev }

\vskip 5 mm
\noi
\centerline{I.E.Tamm Department of Theoretical Physics,
P. N. Lebedev Physical Institute,} \\
\centerline {117924
Leninsky Prospect 53, Moscow, Russia.}

\vskip 3mm \noi

\vskip 9mm

\renewcommand{\theequation}{\arabic{equation}}

\setcounter{equation}{0}

\begin{abstract}                                         
Matrix ge\-ne\-ra\-li\-za\-tions of the N-par\-ti\-cle
quan\-tum-me\-cha\-ni\-cal Ca\-log\-ero mo\-del classifying according to
representations of the symmetric group $S_N$ are considered.
Symmetry properties of the eigen-wave functions in the matrix Calogero models
are analyzed.

\end{abstract}                                           


The Calogero model is the
N-particle quantum-mechanical model on a line with the Hamiltonian
$H =\half
\sum_{i=1}^N \left[ -d_i^2 + x_i^2 \right] + g\sum_{j < i}^N
(x_i-x_j)^{-2}$
where $d_i=\frac \partial {\partial x_i}$.
It gives a prime example of a solvable N-body quantum
mechanical model \cite{calo2,peol}, which has interesting physical
applications. In particular, it is closely related to the matrix models
\cite{ka,ol}, while the generalized differential operators which underly
integrability of the model appear in a number of quite different problems such
as, e.g., the decoupling equations in certain formulations of conformal models
\cite{gose}-\cite{se} and the problem of quantization on the sphere and
hyperboloid \cite{sphy}. It was shown in \cite{lm} - \cite{bhvk} that the
Calogero model can be interpreted as the one-dimensional reduction of
the full anyon problem.
Recently it was argued \cite{gone} that the Calogero
model (and its trigonometric generalization) can be identified with the 2D
Yang-Mills theory on a cylinder. Another intriguing link is that the
higher-spin
gauge theories in three \cite{hs3} and four \cite{hs4} space-time dimensions
exhibit infinite-dimensional symmetries, higher-spin symmetries, described by
the algebra of observables of the Calogero model.

One can speculate that the reason for all these links of the Calogero model
is that the algebraic
structures underlying it are as fundamental as those of the
ordinary harmonic oscillator. The matrix generalizations of the
Calogero model we focus on in this letter may get useful applications as
well.

The spectrum of the Hamiltonian $H$
was found by Calogero \cite{calo2}.
The singularities of $H$ at the
planes $x_i=x_j$ force one to consider the wave functions
of $H$ separately
in $N!$ distinct domains singled out by the sets of inequalities
$x_{i_1}<x_{i_2}<...<x_{i_N}\,$.
Performing the similarity
transformation $ \Psi = \beta^{\nu}\Phi,$ where $\beta=
\prod_{x_i>x_j}(x_i -x_j)\,,$
so that for
$g=\nu(\nu-1)$ the
transformed Hamiltonian $H_{Cal}=\beta^\nu H\beta^{-\nu}$ takes
the
form
\be H_{Cal} =-\half
\sum_{i=1}^N \left[ d_i^2 -x_i^2                
+ \nu \sum_{j\neq i} \frac 2 {x_i - x_j}
d_i \right] \,,
\ee
Calogero argued \cite{calo2}
that, for $\nu>0$,
regular eigenfunctions
$H_{Cal}\Phi_n=E_n\Phi_n$
are of the form
$
\Phi_n=\phi_{nk}(r)P_k(x) \, , \,\, r^2=\frac 1 N
\sum_{i<j}(x_i-x_j)^2$
where
$P_k(x)$ are homogeneous polynomials of degree $k$ obeying
the "generalized harmonic equation"
\be
\left(\sum_{i=1}^N d_i^2+\nu\sum_{i\neq j}\frac
1 {(x_i-x_j)}(d_i-d_j) \right)P_k=0
\ee
while $\phi_{nk}(r)$ obeys cer\-tain equ\-ation
\cite{calo2} which fixes
ener\-gy spect\-rum $E_n\,$.
Ca\-lo\-ge\-ro
proved that every polynomial obeying
(2) is some symmetrical polynomial of $x_i$.
One interpretation of
this elegant result is
that the model automatically selects the subspace of
totally symmetric wave functions which extend to the whole coordinate space.

More
recently, it was shown \cite{brin3} how one can
construct the
set
of eigen-wave functions for the Calogero model, (thus
describing solutions of (2)), with the aid of the approach
based on the permutation operators $K_{ij}$ interchanging the
coordinates $x_i$ and $x_j$. The basic point is that by
introducing
\be
D_i =d_i + \nu \sum_{j\neq i} \frac 1 {(x_i
-x_j) } (1-K_{ij} )
\ee
with $K_{ij}$ obeying the properties
$
K_{ij}x_j=x_iK_{ij} \, ,\, \,\,\,
K_{ij}K_{jl}=K_{il}K_{ij}=K_{jl}K_{il}\,
,\,\,\,$ $ K_{ij}=K_{ji}\, ,\,\,\,   (K_{ij})^2=1\,,
$
one observes, first, \cite{{dunk1},{poly6},{brin3}}
that $[D_i,D_j]=0$ and,
second, that
one can define \cite{brin3} such creation and annihilation operators
$a_i^\mp=\frac{1}{\sqrt 2}(x_i\pm D_i)\,,$ obeying the commutation
relations
\be
[a_i^\pm ,a_j^\pm] = 0 \,,\qquad
 \left[a_i^-
,a_j^+\right]=\delta_{ij} (1+\nu\sum_{l}
K_{il})-\nu K_{ij}=  A_{ij}
\,\, ,
\ee
that the Hamiltonian
$H_{Univ}=\half\sum_i\left\{a_i^+ ,a_i^-\right\}$
fulfills the basic
property
\be
[H_{Univ},a_i^\pm]=\pm a_i^\pm
\ee
and turns out to be related with the original Calogero Hamiltonian
(1) in the following simple way
\be
H_{Univ}=
H_{Cal}+ \half\nu \sum_{j\neq i}
\frac 1 {(x_i -x_j)^2 }(1-K_{ij} )\,.
\ee

Based on
(5) one easily constructs \cite{brin3} the set of
eigen-wave functions of the universal Calogero Hamiltonian
$H_{Univ}$ (6) via the standard procedure
by defining the ground state through $a_i^-|\,0\rangle =0\,,$
$K_{ij}|\,0\rangle =|\,0\rangle\,.$
\footnote
{It should be noted that restriction $\nu>-\frac 1 N$ leads to
the vacuum $|\,0>$ with a finite norm and therefore the restriction
$\nu>0$ guarantees the finiteness of the norm of vacuum for every N.
A detailed consideration of more delicate cases with
non-normalizable vacua and small $\nu$ when some singular
wave functions may also exist will be given
elsewhere.  } 
Since  the second term on the r.h.s.  of (6) trivializes
for totally symmetric states, one observes that for this case $H_{Univ}$
amounts
to $H_{Cal}\,,$
thus recovering the totally symmetric wave functions of the Calogero model.  It
is worth mentioning that the universal Calogero Hamiltonian
$H_{Univ}$ is well defined for the wave functions having arbitrary
symmetry properties while the original
Hamiltonian (1) makes sense only when it coincides with $H_{Univ}$ , i.e. when
the second term on the r.h.s. of (6) trivializes.

The question we address in this paper is whether there
exist other quantum-mechanical models which
amount to the universal
Calogero model (6) for subspaces of
wave functions corresponding to one or another Young
diagram.
We show that such quantum-mechanical models are described by the
Hamiltonians
\be
\hat{H}_{Cal}=H_{Cal}I + \half\nu \sum_{i\neq j}\frac 1
{(x_i-x_j)^2} (I-T_{ij})\,.
\ee
Here $I$ is the unit $m\times m$ matrix
and $T_{ij}=T(p_{ij})$
where $T$ is some
$m\times m$ matrix unitary
representation of the symmetric group with elementary
$i \leftrightarrow j$ permutations
$p_{ij}\,.$
Since $T_{ij}^2=I\,,$ each unitary matrix
$T_{ij}$ is hermitian, that ensures hermiticity
of the Hamiltonian (7).
\footnote
{ To avoid misunderstandings, let us emphasize that,
in contrast to $K_{ij}$,
$m\times m$ $x$-independent matrices $T_{ij}$ commute with $x_i$
and $d_i$.}
It is worth mentioning that both supersymmetric
\cite{{free1},{bhvk}} and matrix \cite{minn1}
generalizations of the
Calogero Hamiltonian considered previously correspond
to particular cases of (7) for some (reducible)
representations $T$ of the symmetric group.

The Calogero result on the symmetry property of the wave functions
extends to the Hamiltonian (7) as follows:
 non-singular eigenfunctions of the
Hamiltonian (7) with $\nu>0$ exist if they obey the
conditions
\be
K_{ij}\Phi=T_{ij}\Phi
\ee
for all $i$ and $j\,$,
i.e. when the Hamiltonians (6) and (7) coincide.

This fact is the main result of this paper which implies in
particular that the condition (8) which was shown in
\cite{{minn1},{bhvk}}
to be convenient to impose to solve the problem is
a sort of necessary condition which cannot be avoided.
Effectively this means that the action of the
symmetric group $S_N$  on the coordinates
$x_i$ generated by $K_{ij}$ realizes the same
representation of $S_N$ as $T_{ij}$ do.
(In particular when
$T_{ij}=1$ the wave functions turn out to be symmetrical.)
The proof of this statement can be given as follows:

\noi
(i) one easily checks that if $\Phi$ (which is $m$-column) is some
solution of
\be
\hat{H}_{Cal}\Phi=E\Phi\,\, ,
\ee
then $K_{ij}T_{ij}\Phi$ and
therefore $Q_{ij}=(1-K_{ij}T_{ij})\Phi$ are some its solution
either;

\noi
(ii) multiplying the both sides of (9)
by $(x_i-x_j)^2$ and setting
$x_i=x_j$ one observes that
$(1-T_{ij})\Phi \mid_{x_i=x_j}=0$.
Since
$K_{ij}$ trivializes for $x_i=x_j$, one concludes that
$Q_{ij}\mid_{x_i=x_j}=0$.

\noi
(iii) using this one then proves along the lines of the original
Calogero proof that $Q_{ij}=0$.
The main steps are as follows.
Suppose that $Q_{ij}\neq 0\,$.
Then $Q_{ij}=(x_i-x_j)^l R_{ij}$
with some positive integer $l$ and the column
$R_{ij}\mid_{x_i=x_j}\neq 0$.
Substituting this back into (9) taking in account (i)
and analyzing the
lowest order terms in $(x_i-x_j)$ one gets that
\be
\left(l(l-1)+2\nu l-\nu (1-T_{ij})
\right)R_{ij}\mid_{x_i=x_j}=0\,.
\ee
Now one observes that $T_{ij}K_{ij}Q_{ij}=-Q_{ij}$.
Therefore $T_{ij}K_{ij}R_{ij}=-(-1)^l R_{ij}$ and
$T_{ij}R_{ij}\mid_{x_i=x_j}$ $=-(-1)^l R_{ij}\mid_{x_i=x_j}$.
Hen\-ce ei\-ther
$(l(l-1)+2\nu l-\nu (1+$ $(-1)^l))=0$
which is  im\-pos\-sible for $l>0$ and $\nu>0$,
or $R_{ij}\mid_{x_i=x_j}=0$, thus com\-ple\-ting the proof.

Thus, entire eigen-wave functions of the Hamiltonian (7)
can exist only when
$Q_{ij}=0$ and therefore (8) is true.
{}From the results of \cite{brin3} it follows that such
solutions of (9) do exist.
Actually, let $T$ be some irreducible
representation of the symmetric group $S_N$ described by
an appropriate Young tableaux.
Let $R(a_i^+)$ be $m$-column of homogeneous polynomials of $a_i^+$
of degree $k$
satisfying the condition $K_{ij}RK_{ij}^{-1}=T_{ij}R$ for
all $i$ and $j$.
It is evident that for every $m$-dimensional representation $T$ of
$S_N$ there exist such sufficiently large $k$
that such a column $R(a_i^+)$ exists. Applied to the symmetric
groundstate of the Hamiltonian (6) it gives some solution of (9):
$\hat{H}_{Cal}R|\,0\rangle =H_{Univ}R|\,0\rangle $ $
=(RH_{Univ}+kR)|\,0\rangle $
$=(E_0+k)R|\,0\rangle $.

The general structure of creation and
annihilation operators
for the Hamiltonian (7) is as follows.
Any annihilation operator $A_n$ is $m\times m$-matrix
operator mapping eigenfunctions of (9) with some eigenvalue $E$
to eigenfunctions having the eigenvalue $E-n$.
Matrices $A_n$ have to preserve (8), i.e.
if a vector-function $\Phi$ belongs to the space of linear combinations
of eigenfunctions of (9) then $K_{ij}A_n\Phi=T_{ij}A_n\Phi$
and hence
the restriction of $A_n$ (which we will identify with $A_n$)
to this space satisfy
the following conditions
\be
K_{ij}A_n K_{ij}^{-1}=T_{ij}A_n T_{ij}^{-1}
\ee
for any $i$ and $j$.
The defining
relation  $[\hat{H}_{Cal},A_n]=-nA_n$
along with (11) leads to
\be
\left[H_{Univ},\, A_n\right]=-nA_n
\ee
Since $H_{Univ}$ is proportional to the unit matrix $I$
the relation (12) is true for all elements of matrix $A_n$
separately and one concludes that n-degree
annihilation operators $A_n$ for the Hamiltonian (9)
are matrices obeying (11) with elements depending on
$a_i$ and $a_i^+$ each having grading $-n\,$.
We hope to present a constructive
description of $A_n$ elsewhere.

One can speculate that the models under investigation describe
interactions of several groups of particles with abnormal
mutual statistics.
For the general case, various types of
interacting matrix Calogero models are classified
according to irreducible representations of the symmetric
group $S_N$.

As a simplest example let us consider
the case corresponding to the Young diagram
with two rows containing $N-1$ boxes and $1$ box respectively.
It is convenient to describe the space of
this representation via column-vectors with the components
\be
\Phi_i=(1-K_{iN})F \,\,\, ,\,\,\, i=1,...,N-1
\ee
where $F$ is some function symmetric under transpositions
of the first $N-1$ variables: $K_{ij}F=F$ for
$i,j=1,2,...,N-1$.

The action of the operators $K_{ij}$ on this columns is
the same as the action of some $(N-1)\times (N-1)$
matrices $\tilde{T}_{ij}$,
$(K_{ij}\Phi)_l=\sum_{k=1}^{N-1}
(\tilde{T}_{ij})_{lk}\Phi_k$:
\be
K_{ij}\Phi_l=\Phi_l\,\,,\,\,\,when \,\,\,\,i,j,l=1,2,...,N-1
\,\,,\,\,\,i\neq l\,\,,\,\, j\neq l
\nonumber \\
K_{ij}\Phi_j=\Phi_i\,\,,\,\,\,when \,\,\,\,i,j,=1,2,...,N-1 \nonumber
\\
K_{iN}\Phi_i=-\Phi_i\,\,,\,\,\, i,=1,2,...,N-1 \nonumber \\
K_{iN}\Phi_j=\Phi_j-\Phi_i\,\,,\,\,\, when
\,\,\,\,i,j,=1,2,...,N-1\,\,,\,\,\,i\neq j
\ee
The matrices $\tilde{T}_{ij}$ are
equivalent to unitary matrices $T_{ij}=Q^{-1}\tilde{T}_{ij}Q$
where $Q$ is any matrix satisfying the conditions
\be
\sum_i (Q_{ki})^2=2 \qquad ,\qquad \sum_i
(Q_{ki}-Q_{li})^2=2\qquad k\neq l
\ee
$i.e.$ its $N-1$ rows together with zero can be interpreted as coordinates
of the apices of some rectilinear $N$-hedron in $(N-1)$-dimensional space.
For example, one can fix $Q_{ij}=\delta_{ij}-\frac 1 {N-1} (1+\sqrt N)$,
$(Q^{-1})_{ij}=\delta_{ij}-\frac 1 {N-1} (1+\frac 1 {\sqrt N})$.

\medskip
{\bf Acknowledgments}

\noindent
This work is supported in part by the Russian Fund of Fundamental Research,
grant N67123016.


\vfill

\end{document}